\documentclass[aps,pra,showpacs,superscriptaddress]{revtex4}
\newcommand{\bb}{\begin{eqnarray}}
\newcommand{\ee}{\end{eqnarray}}

\begin{document}

\title{Fermion pair production in planar Coulomb and
Aharonov--Bohm potentials}

%----  RevTex
\author{V. R. Khalilov}
 \affiliation{Faculty of Physics, Moscow State University, 119899, Moscow,
Russia}
\author{Choon-Lin Ho}
 \affiliation{Department of Physics, Tamkang University, Tamsui
25137, Taiwan, R.O.C.}

%--   LaTex
%\author{V.R. Khalilov$^1$ and Choon-Lin Ho$^2$\footnote{Corresponding author}}
%\date{
%{\small \sl $^1$Faculty of Physics, Moscow State University,
%119899, Moscow, Russia\\ $^2$Department of Physics, Tamkang
%University, Tamsui 25137, Taiwan, R.O.C.}}

%\date{Dec 8, 2008} %{Aug 6, 2008}

%\maketitle
%-----------------------------
\begin{abstract}
Exact analytic solutions are found for the Dirac equation in 2+1
dimensions for a spin-one-half particle in a combination of the
Lorentz 3-vector and scalar Coulomb  as well as Aharonov--Bohm
potentials. We employ the two-component Dirac equation which
contains a new parameter introduced by Hagen to describe the spin
of the spin-1/2 particle. We derive a transcendental equations
that implicitly determine the energy spectrum of an electron near
the negative-energy continuum boundary and the critical charges
for some electron states. Fermion pair production from a vacuum by
a strong Coulomb field in the presence of the magnetic flux tube
of zero radius is considered. It is shown that the presence of the
Ahanorov--Bohm flux tends to stabilize the system.
\end{abstract}

\vskip 1cm


%\noindent{Keywords: Aharonov--Bohm potential, Coulomb potential,
%pair production, critical charge}

\maketitle %for RevTex

%\newpage

\section{Introduction}

The  Aharonov-Bohm (AB) effect is one of the most intriguing
effects of a truly quantal nature \cite{AB}. Ever since its
discovery, the AB effect has been analyzed in various physical
situations in numerous works \cite{AB-review}. In recent years
there has been considerable interest in the problem of scattering
of spin-1/2 particle off an AB potential in 2+1 dimensions. The
results of \cite{AB} for nonrelativistic case modified by using
the Dirac equation in 2+1 dimensions were applied to many
problems. For instance, solutions to the two-component Dirac
equation in the AB potential were first discussed by Alford and
Wilczek in \cite{aw} in a study of the interaction of cosmic
strings with matter. In particular, a mechanism of particle
production due to the nonstatic AB potential of the moving cosmic
string was discussed in \cite{aw}.  Relativistic quantum AB effect
was studied in Ref. \cite{hkh} for the free and bound fermion
states by means of exact analytic solutions of the Dirac equation
in 2+1 dimensions for a combination of AB potential, Lorentz
three-vector and scalar Coulomb potentials.

In \cite{crh} the scattering of spin-polarized fermions in an AB
potential was  considered in 2+1 dimensions. There the particle
spin was introduced into the two-component Dirac equation as a new
parameter. The term including this new parameter appears in the
form of an additional delta-function interaction of spin with
magnetic field in the Dirac equation. Solutions of the Dirac
equation were then interpreted for the case of 3+1 dimensions.
Similar problems were also discussed in \cite{KhaHo07} by taking
the AB flux tube with a small but finite radius in
(3+1)-dimensional Pauli equation, and in (2+1)-dimensional Dirac
equation.

In this paper we would like to study  how various physical fields
affect the properties of a bound Dirac fermion in 2+1 dimensions.
Specifically, we study the energy spectrum of the fermion in a
combination of the Lorentz 3-vector and scalar Coulomb as well as
AB potentials. We also consider the influence of the magnetic flux
tube of zero radius on the so-called critical charge which
determines the onset of instability of the system, and on the
probability of production of an electron-positron pair from a
vacuum by a strong vector Coulomb field.  We note here that in 3+1
dimensions, analytic solutions of such problems, even for the
Schr\"odinger equation in the Coulomb and AB potentials, have not
yet been found.

This paper is organized as follows. In section II we find the
exact bound states solutions of the Dirac equation in 2+1
dimensions for a combination of the Lorentz 3-vector, scalar
Coulomb and AB potentials for the spin-1/2 particle. The formulas
for eigen-energies of the relativistic fermion are obtained and
discussed. In section III the critical charge and the probability
of production of an electron-positron pair from a vacuum by a
strong vector Coulomb field are calculated. Using a simplified
model with  a truncated Coulomb potential we show that the
critical charge and the production probability of pair are
influenced by the magnetic flux and spin particle.

\section{Energy spectrum of a Dirac fermion in
planar Coulomb and Aharonov--Bohm potentials}

The (2+1)-dimensional Dirac equation of a fermion of mass $m$ and
charge $e=-e_0<0$ in a vector potential $A_{\mu}$ and a Lorentz
scalar potential $U$ is ($c=\hbar=1$)
 \bb
 (\gamma^{\mu}
P_{\mu} - m-U)\Psi = 0, \label{Dirac}
 \ee
 where $P_\mu = -i\partial_{\mu} - eA_{\mu}$ is the
generalized fermion momentum operator. The Dirac $\gamma^{\mu}$
matrices are conveniently defined in terms of the Pauli spin
matrices as
 \bb
 \gamma^0= \sigma_3,\quad
\gamma^1=is\sigma_1,\quad \gamma^2=i\sigma_2, \label{1spin}
 \ee
Following \cite{crh}, here $s$ is a new parameter characterizing
twice the spin value $s=\pm 1$ for spin ``up" and ``down",
respectively.

We are interested in finding exact analytic solutions of the Dirac
equation for both signs of $s$ in  an AB potential, which is
specified in the Cartesian or cylindrical coordinates as
 \bb
A^0=0,\quad A_x=-\frac{By}{r^2},\quad A_y=\frac{Bx}{r^2}; \quad
A^0=0,\quad A_r=0,\quad A_{\varphi}=\frac{B}{r},
\nonumber\\
\quad r=\sqrt{x^2+y^2}, \quad
\varphi=\tan^{-1}(y/x)\phantom{mmmmmmmmm} \label{eight}
 \ee
  and a
Lorentz $3$-vector potential ($A^{\mu}(r)$) and a scalar
potential($U(r)$) potential defined by
 \bb A^0(r) =
\frac{a}{|e|r},\quad A_r=0, \quad A_{\varphi}=0; \quad
U(r)=-\frac{b}{r} \quad (a,b>0) \label{super}
 \ee
In \cite{hkh} only the case for $s=1$ was considered. Note that if
$e_0B=N$, where $N$ is an integer, then the magnetic field flux is
quantized as $\Phi=\Phi_0 N$, where $\Phi_0\equiv 2\pi/e_0$ is the
elementary magnetic flux called the ``fluxon''.

The Dirac Hamiltonian for this system is
 \bb
 H_D=\sigma_1P_2-s\sigma_2P_1+\sigma_3(m+U(r))-e_0A_0(r).\label{diham}
 \ee
The vector potential $A_\mu$ in the generalized momentum $P_\mu$
is the sum of the AB potential (\ref{eight}) and the Lorentz
$3$-vector potential (\ref{super}).  The total angular momentum
$J_z\equiv L_z+ s\sigma_3/2$, where $L_z\equiv
-i\partial/\partial\varphi$, is a conserved quantity.

We seek positive energy solutions of Eq.~(\ref{Dirac}) in the form
(see, also, Ref. \cite{HoKha00,KhaHo98,ktmp1})
  \bb
 \Psi(t,{\bf x}) = \frac{1}{\sqrt{2\pi}}\exp(-iEt+il\varphi)
\psi(r, \varphi)~, \label{three}
 \ee
  where $E\geq 0$ is the fermion
energy, $l$ is an integer, and $\psi(r, \varphi)$ is a
two-component function ({\it i.e.} a  $2$-spinor)
 \bb
  \psi(r,
\varphi) = \left( \begin{array}{c}
f(r)\\
g(r)e^{is\varphi}
\end{array}\right).
\label{four}
 \ee
The wave function $\Psi$ is an eigenfunction of the conserved
total angular momentum $J_z$ with eigenvalue $j=l+s/2$.

From the Dirac equation one finds that $f(r)$ and $g(r)$
 satisfy the following  equations
:
 \bb s{df\over dr}-{l+e_0B\over
r}f+\left(E+m+\frac{a-b}{r} \right)g = 0,
\nonumber \\
s{dg\over dr}+{l+s+e_0B\over r}g-\left(E-m+\frac{a+b}{r} \right)f
= 0. \label{eq7}
 \ee
Compared with the situation studied in \cite{KhaHo98} in which
only the vector Coulomb potential ($a$) is present, one notes that
the effect of the AB potential ($B$) appears only in modifying the
angular momentum $l$.  As in \cite{KhaHo98}, we now assume $f(r)$
and $g(r)$ to have the form (see, for example, \cite{blp})
 \bb
 f(r) = \sqrt{m+E}e^{-x/2}x^{\gamma_s}(Q^s_1 + Q^s_2)~, \nonumber \\
 g(r) = \sqrt{m-E}e^{-x/2}x^{\gamma_s}(Q^s_1 - Q^s_2)~,
\label{sys1}
 \ee
  where
 \bb
 x = 2\lambda r,\quad \lambda = \sqrt{m^2-E^2},
\label{notion}
 \ee
 and
  \bb
\gamma_s=-\frac12\pm\sqrt{\left(l+e_0B+\frac{s}{2}\right)^2-a^2+b^2}
\label{eq9}
 \ee
  determines the asymptotic behavior of the wave
function for small $r$.

From Eq.~(\ref{eq9}) one sees that when $a^2<(l+e_0B+s/2)^2+b^2$
the quantity $\gamma_s$ is real, and must be chosen positive to
ensure normalizability of the wave function. If
$a^2>(l+e_0B+s/2)^2+b^2$ then the two roots of $\gamma_s$ are
imaginary and the corresponding wave functions oscillate as $r\to
0$, which indicates the occurrence of Klein's paradox
\cite{bjdl,blp}.  We shall consider this situation in the next
section.  In what follows we shall take
  \bb
\gamma_s=-\frac12+\sqrt{\left(l+e_0B+\frac{s}{2}\right)^2-a^2+b^2}.
\label{eq9a}
 \ee
For wave functions that are finite at $x=0$, the functions
$Q^s_{1,2}$ for $s=\pm 1$ are given by the confluent
hypergeometric function $F(a_1,c_1;x)$:
 \bb
 Q_{1}^s = AF\left(\gamma_s+1-\frac{s}{2} - \frac{aE+mb}{\lambda},
 2\gamma_s+2; x\right), \nonumber
\\  Q_2^s = CF\left(\gamma_s+1+\frac{s}{2} - \frac{aE+mb}{\lambda},
2\gamma_s+2; x\right).
\label{twfour}
 \ee
 The constants $A$ and $C$ are related by \bb C
=
\frac{(s\gamma_s+s/2)-(Ea+mb)/\lambda}{l+e_0B+s/2+(ma+bE)/\lambda}A.
\label{conn}
 \ee

The wave function is normalizable if both the hypergeometric
functions $Q_1^s$ and $Q_2^s$ are reduced to polynomials.  For
this the parameter $a$ of $F(a, c; x)$ must be a negative integer
or zero. Denoting
 \bb
  \gamma_s+1-
\frac{s}{2}-\frac{Ea+mb}{\lambda} = -n_r, \label{deno1}
 \ee
  one
can see that  if $n_r=1, 2, 3,\ldots$, then $Q_1^1$ and $Q_2^1$
are reduced to polynomials for $s=1$. If $n_r=0$, then only
$Q_1^1$ is reduced to a polynomial. But the relation $n_r=0$
implies that
 \bb \gamma_1 + \frac12=\frac{Ea+mb}{\lambda}>0.
\label{deno2} \ee
Then it follows from
 (\ref{deno2}) and the relation
 \bb
\left(\gamma_s +\frac12\right)^2 - \frac{a^2-b^2}{\lambda^2}E^2=
\left(l+e_0B+\frac{s}{2}\right)^2 - \frac{a^2-b^2}{\lambda^2}m^2
\label{1rel} \ee
 that
  \bb
\frac{Eb+ma}{\lambda}=\left|\,l+e_0B+\frac12\,\right|.
\label{deno3}
 \ee
If $l+e_0B+1/2>0$, then $C=0$, hence $Q_2^1=0$ and the required
condition is not violated ($Q_1^1$ is a polynomial). If
$l+e_0B+1/2<0$, then $C=A$, and $Q_2^1$ is still a divergent
function. The following values of $n_r$ are hence admissible: $0,
1, 2, \ldots$ for $l+e_0B+1/2>0$,  and $1, 2, \ldots$ for
$l+e_0B+1/2<0$.

For $s=-1$,  $Q_1^{-1}$ and $Q_2^{-1}$ are reduced to polynomials
if $n_r=0, 1, 2, 3, \ldots$.  But in this case $n_r$ also can be
equal to $-1$.  If $n_r=-1$, then only $Q_2^{-1}$ is reduced to a
polynomial. But for $n_r=-1$ from (\ref{deno1}) one obtains
 \bb
\gamma_{-1} + \frac12=\frac{Ea+mb}{\lambda}>0 \label{deno-2}
 \ee
 and Eq.~(\ref{deno3}) becomes
  \bb
\frac{Eb+ma}{\lambda}=\left|\,l+e_0B-\frac12\,\right|.
\label{deno-3}
 \ee
It is convenient to rewrite the relation (\ref{conn}) for $s=-1$
as
 \bb
  A =-\frac{l+e_0B-1/2+(ma+bE)/\lambda}{(\gamma_{-1}+1/2)+(Ea+mb)/\lambda}C.
\label{conn-}
 \ee
  So, if $l+e_0B-1/2<0$, then $A=0$, hence
$Q_1^{-1}=0$ and the required condition is not violated
($Q_2^{-1}$ is a polynomial). If $l+e_0B-1/2>0$, then $A=-C$, and
$Q_1^{-1}$ is divergent. Hence the following values of $n_r$ are
admissible for $s=-1$: $-1, 0, 1, 2, \ldots$ for $l+e_0B-1/2<0$,
and $0, 1, 2, \ldots$ for $l+e_0B-1/2>0$.

We can rewrite the equation (\ref{deno1}) for the energy spectrum
as
 \bb
  \frac{Ea+mb}{\lambda}=n_r+\gamma_s+1-
\frac{s}{2}\equiv u, \label{spect}
 \ee
 from which we obtain finally the discrete
fermion energy levels in the form
 \bb
 \frac{E_n}{m} =
\sqrt{\left(\frac{ab}{u^2+a^2}\right)^2+ \frac{u^2-b^2}{u^2+a^2}}
-\frac{ab}{u^2+a^2}~. \label{spectrum}
 \ee

One sees that the energy spectrum is influenced by the magnetic
flux through $\gamma_s$ given by Eq.~(\ref{eq9a}). On the other
hand, for the flux that is integer in the unit $\Phi_0$ the energy
spectrum is the same as in the absence of magnetic flux. The
spectrum of the system changes only for flux that is not integer
in the unit $\Phi_0$. For such magnetic fluxes the energy spectrum
is likely to be observed just as the AB effect. For flux that is
not integer or half integer all the energy levels are doubly
degenerate; the levels with $l, n_r+1$, $s=+1$ and $l+1,n_r$,
$s=-1$ coincide. This reflects the fact that the fermion energy
does not depend upon spin in the field configuration considered.

If the scalar Coulomb potential is absent, then the energy
spectrum is given by
 \bb
 E_{n,l} = m\left[1 + \frac{a^2}{(n_r + 1/2- s/2
 + \sqrt{(l+e_0B+s/2)^2-a^2})^2}
\right]^{-1/2}~. \label{twsix}
 \ee
  This expression makes sense only
when $|l+e_0B+s/2|>a$, a condition that forbids the existence of
the  energy levels with $l+e_0B+s/2=0$.

In the nonrelativistic Schr\"odinger limit, the expression for the energy
spectrum becomes
\bb
 E_{non} = - \frac{a^2}{2(n_r + 1/2 -s/2 + |l+e_0B+s/2|)^2}~.
\label{nonr}
 \ee
 For flux $e_0B$ that is not integer or half
integer it has a full analogy with the Rydberg correction.  Note
that for flux that is half integer in the unit $\Phi_0$ the energy
spectrum (\ref{nonr}) depends only on integer number. It can be
observed using spectroscopy. It is interesting that for such
magnetic fluxes  the cross section in the AB scattering is
maximal. It is seen that the eigen-energies of a fermion in these
electromagnetic field combination are periodic function of the
flux like the case of the motion of fermion in a closed ring in
the absence of the two-dimensional Coulomb potential. The energy
spectrum (\ref{nonr}) repeats itself every time when the change in
the flux $e_0B$ is integral.

\section{Critical charge and Pair production in vector Coulomb and
Aharonov-Bohm potential}

We now consider the problem of stability of the system considered
in the previous section when $a$ becomes large.  For simplicity we
ignore the scalar Coulomb potential in this section, i.e. we set
$b=0$.  Such consideration is relevant to the stability of the
vacuum of quantum electrodynamics in a strong Coulomb field. In
the absence also of the AB potential this problem in $3+1$
dimensions had been extensively studied in
\cite{wpw,sgy,yzvp,mig,rfk,smg,cow}. The corresponding system in
$2+1$ dimensions was considered in \cite{KhaHo98,ktmp1}. Now we
would like to see how the presence of the AB potential may affect
the stability of the system.

From Eq.~(\ref{spectrum}) we see that the lowest electron states
in the vector Coulomb and AB potentials are those with $n_r=l=0,
s=1$ and $n_r=-1, l=0, s=-1$. The electron energy in these states
can be written as
 \bb
 E_s = m\sqrt{\frac{(e_0B+s/2)^2-a^2}{(e_0B+s/2)^2}}~. \label{grste}
  \ee
For definiteness, in this section we shall consider positive flux
($e_0B>0$). The case with negative flux can be discussed similarly
with the signs of $l$ and $s$ flipped: it is just the mirror image
of the case with positive flux with respective to the $xy$-plane.
For magnetic flux which is positive and half integer in $\Phi_0$,
the energy of the state with $n_r=-1, l=0, s=-1$ is divergent and
imaginary for any value of $a$.  So it is reliable to assume that
the electron ground state in this case is the state with $n_r=l=0$
and $s=1$ .

The energy $E_0(a)$ of this state as a function of $a$ becomes
zero at $a=a_{0}(B)\equiv e_0B+1/2$ and purely imaginary when
$a>a_{cr}(B)$.  This implies that $E_0(a)$ becomes meaningless at
$a\ge a_{0}(B)$,  and the wave function oscillates with infinite
frequency at the origin as mentioned in Sect.~2.  It may appear
that one cannot determine the spectrum beyond $a_0(B)$. However,
one notes that in reality the source of the Coulomb field has
finite extension so that the potential remains finite at the
origin.  In this case the wave function is regular at the origin
and the energy levels can be traced continuously beyond $a_0(B)$.

The problem of a $(2+1)$-dimensional Dirac particle in a strong
field of a truncated (at small distances) Coulomb potential in the
absence of AB potential was considered in \cite{KhaHo98}, and the
expression for the electron energy spectrum was obtained in
\cite{ktmp1}. There it was shown that as the strength of the
Coulomb source $a$ increases, the lowest energy level (c.f.
Eq.~(\ref{grste})) $E_{s=1}(B=0)=m\sqrt{1-4a^2}$ was pulled
towards the negative continuum. This energy becomes negative for
$a^2>1/4$ (in the truncated potential) and may reach the
negative-energy continuum boundary $-m$. When $E_{s=1}(B=0)$ dives
into the negative continuum the vacuum of quantum electrodynamics
becomes unstable and particle-antiparticle pair is created
spontaneously. The value $a=a_{cr}$ for which the lowest energy
level coincides with $-m$ is called the critical charge for the
ground state.

Let us consider the effect of the AB potential on the stability of
the system qualitatively.  From Eq.~(\ref{eq7}) we see that the
effect of the AB potential appears only in increasing the angular
momentum $l$.  Classically it increases the centrifugal force on
the particle.  Thus it tends to counteract the tendency of the
particle being pulled towards the negative continuum.  Hence one
expects that the AB potential (\ref{eight}) will stabilize the
system against pair production.

Quantitatively, to determine the critical charge $a_{cr}(B)$ in
the presence of the AB potential, it is sufficient to consider the
range of electron energies near the negative-energy continuum
limit $-m$. Introducing functions $F(r)=rf(r)$ and $G(r)=rg(r)$,
and eliminating $G(r)$ from (\ref{eq7}), we obtain the equation
for the function $F(r)$ with $E\approx -m$ in the form \bb
 {d^2F\over
dr^2}+\left(E^2-m^2+\frac{2Ea}{r}+\frac{a^2-(l+e_0B)(l+e_0B+1)}
{r^2}\right)F=0.
 \label{eqlim}
  \ee
The solution for $G(r)$ near $E=-m$ can be found from the equation
\bb
 G(r)=\frac{l+e_0B+1}{a}F(r)-s\frac{r}{a}{dF\over dr}.
\label{conFG}
 \ee

The solution of Eq. (\ref{eqlim}) which tends to zero as  $r\to
\infty$ can be expressed through the Whittaker function of the
form
 \bb
F(r)\sim W_{\beta,i\theta}(2\lambda r),\label{whitt}\ee where \bb
\beta=\frac{Ea}{\lambda},\quad \theta=\sqrt{a^2-(l+e_0B+1/2)^2}
\label{notwhi}
 \ee
 near $E=-m$, or through the MacDonald function of
imaginary order
 \bb
F(r)\sim \sqrt{r}K_{2i\theta}(\sqrt{8mar})\label{macd} \ee
 for $E=-m$.
It follows from Eqs.~(\ref{whitt}) or (\ref{macd}) that the bound
electron state with $E\approx -m$ is localized in space. Such
behavior of the electron state can be easily explained if we treat
Eq. (\ref{eqlim}) as the Schr\"odinger equation with the effective
energy $\epsilon=(E^2-m^2)/2m$ and the effective potential
 \bb
U_{eff}(r)=-\frac{2Ea}{mr}-\frac{a^2-(l+e_0B)(l+e_0B+s)}{2mr^2}.
\label{eff1}
\ee
 In the case $E\approx -m$ for the ground electron
state $l=0$ the potential $U_{eff}(r)$ has a form of a wide
barrier. One notes that at large distances from the Coulomb center
for an electron with energy $E\approx -m$ the effective potential
is not attracting but repulsive. In the presence of the magnetic
flux the height and width of the effective potential barrier
increase for $s=1$ and $e_0B>1$. Therefore, the probability of
pair production by the Coulomb field  decreases in the presence of
the magnetic flux.

For a weak magnetic flux the form of vector Coulomb potential for
$r<R$ is not essential to the principal result.  The calculation
is most easily performed if we consider the simplest model with
 only the 3-vector Coulomb potential
  \bb
   A^0(r) =\frac{a}{|e|R},
\quad A_r=0, \quad A_{\varphi}=0 \label{1super}
 \ee
   in the range
$r\le R$. Then the radial solution $F(r)$ that is finite at $r=0$
in the range $r\le R$ is expressed via the Bessel function of
integer order $|l|$ as
 \bb F(r)\sim rJ_{|l|}(cr),\label{bess}
 \ee
where
 \bb c=\sqrt{\left(E+\frac{a}{R}\right)^2-m^2}. \label{argbes}
 \ee

Applying the continuity relations
 \bb
\left(\frac{G(r)}{F(r)}\right)_{r=R-0}=
\left(\frac{G(r)}{F(r)}\right)_{r=R+0}\label{1cont}
 \ee
 and
taking into account the fact that the parameter $R$ must be small
compared with $1/m$ and also that $E\approx -m$, we obtain the
transcendental equations (for $l=0$) that implicitly  determine
the energy spectrum of an electron near the negative-energy
continuum boundary $-m$ as
 \bb
cR\frac{J_1(cR)}{J_0(cR)}=1-\left(x
\frac{W_{\beta,i\theta}^{\prime}(x)}
{W_{\beta,i\theta}(x)}\right)_{x=2\lambda R},\label{exsst}
 \ee
 and the critical charge
for the ground state
 \bb
2a_{cr}(B)\frac{J_1(a_{cr}(B))}{J_0(a_{cr}(B))}=1-\left(z\frac{K_{i\nu}^{\prime}(z)}
{K_{i\nu}(z)}\right)_{z=\sqrt{8ma_{cr}(B)R}}.\label{grst}
 \ee
Here
 \bb
  \beta=-\frac{ma}{\lambda},\quad
\theta=\sqrt{a^2-(e_0B+1/2)^2},\quad
\nu=2\sqrt{a_{cr}^2(B)-(e_0B+1/2)^2},\label{notcr}
 \ee
  and the prime denotes differentiation with
respect to the argument $x$ of the Whittaker function in
(\ref{exsst}), or the argument $z$ of the MacDonald function in
(\ref{grst}).

Equations (\ref{exsst}) and (\ref{grst}) can only be solved
numerically.  These equations can be simplified somewhat for
$Rm\ll 1$.  For instance, to simplify Eq. (\ref{grst}),  we
represent the MacDonald function in the form \cite{GR}
 \bb
K_{\mu}(z)=\frac{\pi}{2\sin(\mu\pi)}[I_{-\mu}-I_{\mu}],
\label{mac}
\ee
where
 \bb
I_{\mu}=\sum\limits_{k=0}^{\infty}\frac{1}{k!\Gamma(\mu+k+1)}
\left(\frac{z}{2}\right)^{\mu+2k}. \label{imbes}
 \ee
 Keeping only the lowest order terms in the
expansion of the MacDonald function at small values of the
argument, we obtain
 \bb
K_{i\nu}(z)\sim
-\left(\frac{\pi}{\nu\sinh(\nu\pi)}\right)^{1/2}\sin\left(\nu\ln(|z|/2)
+\arg\Gamma(1-i\nu)\right),
\label{mac1}
 \ee
where $\Gamma(z)$ is the Euler gamma function. With
Eq.~(\ref{mac1}), we finally obtain an approximate form of Eq.
(\ref{grst}) at small $Rm$:
 \bb
2a_{crb}\frac{J_1(a_{crb})}{J_0(a_{crb})}
=1-\nu\cot\left[\nu\ln(|z|/2)+\arg\Gamma(1-i\nu)\right].
\label{grsts}
 \ee
 Here $a_{crb}\equiv a_{cr}(B)$.

Numerical solution of Eq. (\ref{grsts})  gives $a_{crb}\approx
0.79$ for $B=0$, $a_{crb}\approx 0.84$ for $e_0B=0.1$,
$a_{crb}\approx 0.91$ for $e_0B=0.2$ at $Rm=0.02$, and
$a_{crb}\approx 0.70$ for $B=0$, $a_{crb}\approx 0.77$ for
$e_0B=0.1$, $a_{crb}\approx 0.84$ for $e_0B=0.2$ at $Rm=0.006$.
One sees that the critical charge increases with the increase of
$B$ and decreases with the decrease of $Rm$. Therefore, the
instability of the vacuum of quantum electrodynamics in a strong
vector Coulomb field in 2+1 dimensions in the presence of the
magnetic flux tube of a very small radius must occur at larger
critical charge, as compared to that in the absence of the
magnetic flux. Thus, the magnetic flux stabilizes the vacuum of
quantum electrodynamics in the vector Coulomb field.

Similarly, Eq. (\ref{exsst}) can be simplified. With the condition
$Rm\ll 1$ we use the representation of the Whittaker function for
small values of the argument in the form \bb
 W_{\beta,i\theta}(x)
&=&\frac{\Gamma(2i\theta)}{\Gamma(1/2-\beta+i\theta)}x^{1/2-i\theta}+
\frac{\Gamma(-2i\theta)}{\Gamma(1/2-\beta-i\theta)}x^{1/2+i\theta},
\label{exswh}\\
&=&\frac{|\Gamma(2i\theta)|}{|\Gamma(1/2-\beta+i\theta)|}2x^{1/2}\cos(\Phi(x)),
 \label{exswh1}
 \ee
where
  \bb \Phi(x) = -\theta\ln x +
 \arg\Gamma(2i\theta)-\arg\Gamma(1/2-\beta+i\theta).
 \label{exswh2}
 \ee
Taking the derivative with respect to $x$ in Eq. (\ref{exswh1})
and substituting the resulting expression and function
(\ref{exswh1}) in Eq. (\ref{exsst}), we finally obtain the
simplified transcendental equation
 \bb
  -\theta\ln(2\lambda R) +
 \arg\Gamma(2i\theta)-\arg\Gamma(1/2-\beta+i\theta) = \tan^{-1} Y
 +\pi n_r,  \label{exswh4}
 \ee
  where
 \bb
Y=\theta^{-1}\left(\frac{1}{2}-cR\frac{J_1(cR)}{J_0(cR)}\right).
\label{phase}
 \ee
  The equation for the energy spectrum for any
integer $l$ can be derived by the analogous way.

Equation (\ref{exswh4}) with $a<a_{crb}$ implicitly determines the
eigen spectrum of bound electron states for $l=0$ with $m>E>-m$.
It can be shown there are real solutions of this equation only for
$a<a_{crb}$, but with $a>a_{crb}$ there is a formal solution of
the form $E=E_0-iw$, where $E_0=-m-c_1(a-a_{crb}),\quad c_1\sim 1$
for the lowest state. With $a-a_{crb}\ll a_{crb}$, the imaginary
part $w$ is exponentially small. Such a solution can be found from
formula (\ref{exswh4}) through analytic continuation of $E$ as a
function of $a$ into the range $a>a_{crb}$. However, for
$a-a_{crb}\ll a_{crb}$, the imaginary part can be more readily
determined in another way. Indeed, the appearance of the imaginary
part means that for $a>a_{crb}$, the corresponding Dirac equation
has only a formal solution with $E=E_0-iw$ for the electron
states. However, for $a>a_{crb}$, the same equation also describes
a positron with the energy $E_0=m+c_1(a-a_{crb})$ since the Dirac
equation for a positron for $a>a_{crb}$ can be obtained from the
Dirac equation for an electron by replacing $E$ with $-E$ and $a$
with $-a$. Hence, for $a>a_{crb}$, the positron states are
quasi-stationary. For $a-a_{crb}\equiv \Delta a \ll a_{crb}$, the
width of the quasi-stationary level $w$ can be estimated in the
semiclassical approximation. To obtain this estimate, we must
compute the transmission coefficient through the potential barrier
in Eq. (\ref{eqlim}). We note that the width $w$ is half of the
reciprocal of the positron lifetime or twice the probability of
pair creation by the Coulomb field. If $\Delta a \ll a_{crb}$, the
positrons created by the field are very slow, and the Coulomb
barrier is hardly transparent for them. The probability of pair
creation is therefore exponentially small, i.e.,
 \bb w\sim
m\exp\left[-2\pi a\left(\frac{m}{\sqrt{E^2-m^2}}-1 \right)\right]
\cong m\exp\left(-c_2\sqrt{\frac{a_{crb}}{\Delta a}}\right), \quad
c_2\sim 1. \label{create}
\ee

\section{Summary}

In this paper we solve exactly the Dirac equation in 2+1
dimensions in a combination of  Lorentz 3-vector, scalar Coulomb
and AB potentials. We employ the two-component Dirac equation
which contains a new parameter introduced by Hagen to describe the
spin of the spin-1/2 particle.  We study the energy spectrum of an
electron near the negative-energy continuum boundary and the
critical charges for some electron states. Fermion pair production
from a vacuum by a strong Coulomb field in the presence of the
magnetic flux tube of zero radius is also considered. It is shown
that the presence of the AB flux tends to stabilize the system.

Finally, we note that solutions to the two-component Dirac
equation in the AB potential coincide with the solutions of the
Dirac equation in 2+1 dimensions for a massive neutral fermion
with the anomalous magnetic moment in a point charge placed at the
origin $z=0$ (see, for example, Ref.\cite{ktmp}). In the
three-dimensional space, such a field corresponds to the electric
filed of a thin thread that is perpendicular to the plane $z=0$
and carries  the electric charge with the constant linear density.
Thus, the solutions to the two-component Dirac equation in the AB
potential can be directly applied to the planar scattering  of a
massive neutral fermion with anomalous magnetic moment interacting
in the electric field of the thin thread, which was first
predicted by Aharonov and Casher in Ref. \cite{ahc}.

 \vspace{1cm}
\begin{acknowledgments}
%\vskip 2cm \centerline{\bf Acknowledgments}
 This work was
supported  by a Joint Research Project of the National Science
Council (NSC) (Taiwan) under Grant No. NSC 95-2911-M-032-001-MY2
(C.L.H), and the Russian Foundation for Basic Research (RFBR)
under Grant No. NSC-a-89500.2006.2 and NSC-RFBR No. 95WFD0400022
(contract No. RP06N04) (V.R.K), and in part by the Program for
Leading Russian Scientific Schools (Grant No.
NSh-5332.2006.2)(V.R.K.). V.R.K. thanks  Dr. I. Mamsurov and A.
Borisov for useful discussions. Part of the work was completed
during C.L.H.'s visit to the Faculty of Physics at the Moscow
State University.
\end{acknowledgments}

%\vspace{1cm}
%\newpage

\end{document}